\documentclass{emulateapj}

\shorttitle{Ultra-massive fast-spinning white dwarf}

 \shortauthors{S.~Mereghetti~et~ al.}

\def\approxgt{\mathrel{\hbox{\rlap{\lower.55ex \hbox {$\sim$}}
        \kern-.3em \raise.4ex \hbox{$>$}}}}
\def\approxlt{\mathrel{\hbox{\rlap{\lower.55ex \hbox {$\sim$}}
        \kern-.3em \raise.4ex \hbox{$<$}}}}
\def \xmm {\emph{XMM-Newton} }

\def\pdot {\dot P}

\def\ltsima{$\; \buildrel < \over \sim \;$}
\def\lsim{\lower.5ex\hbox{\ltsima}}
\def\gtsima{$\; \buildrel > \over \sim \;$}
\def\gsim{\lower.5ex\hbox{\gtsima}}

\def\msun{~M_{\odot}}
\def\lsun{~L_{\odot}}
\def\rsun{~R_{\odot}}
\def\Mdot {\dot M}
\def\hd {HD 49798}
\def\rx {RX J0648.0--4418}
\def\hr {HD\,49798/RX\,J0648.0--4418}

\begin{document}


\title{X-ray and optical observations of the unique binary system    \hr\  }

\author{S. Mereghetti\altaffilmark{1}, N. La Palombara\altaffilmark{1},
A. Tiengo\altaffilmark{1}, F. Pizzolato\altaffilmark{1},
P. Esposito\altaffilmark{2}, P. A. Woudt\altaffilmark{3},
G.~L. Israel\altaffilmark{4}, L. Stella\altaffilmark{4}}
 \affil{$^{1}$ INAF - Istituto di Astrofisica Spaziale e Fisica
Cosmica Milano, via E.~Bassini 15, I-20133 Milano, Italy; \url{sandro@iasf-milano.inaf.it}}
 \affil{$^{2}$  INAF - Osservatorio Astronomico di Cagliari, localit\'a Poggio dei Pini, strada 54,
 I-09012 Capoterra, Italy}
 \affil{$^{3}$ Astronomy Department, and Astrophysics, Cosmology and Gravity Centre, University of Cape Town, Private Bag X3, Rondebosch 7701, South Africa}
 \affil{$^{4}$ INAF - Osservatorio Astronomico di Roma, via Frascati 33, I-00040 Monteporzio Catone, Italy}

\begin{abstract}

We report the results of \xmm\ observations of \hr , the only
known X-ray binary consisting of a hot sub-dwarf and a white
dwarf. The white dwarf rotates very rapidly (P=13.2 s) and has
a dynamically measured mass of 1.28$\pm$0.05$\msun$. Its X-ray
emission consists of  a strongly pulsed, soft component, well
fit by a blackbody with kT$_{BB}\sim$40 eV, accounting for most
of the luminosity, and a fainter hard power-law component
(photon index $\sim$1.6). A luminosity of $\sim$10$^{32}$ erg
s$^{-1}$ is produced by accretion onto the white dwarf of the
helium-rich matter from the wind of the companion, which is one
of the few hot sub-dwarfs showing evidence of  mass-loss. A
search for optical pulsations at the South African Astronomical
Observatory  1.9-m telescope gave negative results. X-rays were
detected also during  the white dwarf  eclipse. This emission,
with luminosity 2$\times$10$^{30}$ erg s$^{-1}$, can be
attributed to \hd\ and represents the first detection of a hot
sub-dwarf star in the X-ray band. \hr\ is a post-common
envelope binary which most likely originated from a pair of
stars with masses $\sim$8--10 $\msun$.  After the current
He-burning phase,  \hd\ will expand and reach the Roche-lobe,
causing a higher accretion rate onto the white dwarf which can
reach the Chandrasekhar limit. Considering the fast spin of the
white dwarf, this could lead to the formation of a millisecond
pulsar. Alternatively, this system could be a Type Ia supernova
progenitor with the appealing characteristic of a short time
delay, being the descendent of relatively massive stars.

\end{abstract}


\section{Introduction}

Interacting binaries containing accreting white dwarfs
constitute a large fraction of the population of bright
Galactic X-ray emitters. These systems, including both
persistent and transient sources, display a rich variety of
properties, determined mainly by the strength of the white
dwarf's magnetic field  and the nature of the mass donor star
(see, e.g., \citet{kuu06}). Most accreting white dwarfs are
found in cataclysmic variables, where accretion proceeds
through Roche-lobe overflow.  The mass donor in these systems
is typically a main sequence star or another white dwarf, but
also sub-giant and giant companion stars have been observed.
Symbiotic systems, where a white dwarf accretes from the
stellar wind of a massive supergiant, have also been detected
as X-ray sources.

The system discussed here is the only known X-ray binary
composed by a white dwarf (\rx ) and a hot sub-dwarf star (\hd
). Hot sub-dwarfs are evolved low mass stars that lost most of
their hydrogen envelopes and are   believed to be in the helium
core burning stage (for a recent review on hot sub-dwarfs see
\citet{heb09}). They are spectroscopically classified in
different types: sdB (with effective temperature
T$\sim$25,000--28,000 K), sdOB (with T$\sim$33,000--40,000 K),
and sdO (with T$>$40,000 K). One possible mechanism responsible
for the loss of their hydrogen envelopes is non-conservative
mass transfer in a binary \citep{pod04}.

Being one of the brightest hot sub-dwarfs (apparent magnitude V=8.3),
\hd\ has been the object of several studies in the optical/UV wavebands.
The first spectroscopic observations led to its classification as a sub-dwarf of O6 spectral
type \citep{jas63} and showed radial velocity variations, later found to be caused by orbital
motion with a  period of  1.55 days \citep{tha70,sti94}.
No spectroscopic evidence for a secondary could be found, leading to the early suggestion that
the companion could be a white dwarf \citep{tha70}.
Soft X-ray emission from the direction of \hd\ was first
detected with the \textit{Einstein Observatory} in 1979,
but it could be studied in   detail only with a   \textit{ROSAT} satellite observation performed in 1992.
The \textit{ROSAT} data showed a very soft spectrum and led to the discovery
of a strong modulation with a period of 13.2 s \citep{isr97}.
While the periodicity clearly indicated the presence of a compact object,
the poorly constrained spectral fit resulted in a large uncertainty in the X-ray luminosity,
making it impossible to distinguish between a neutron star and a white dwarf.

Thanks to a long observation carried out with  the \xmm
satellite  in May 2008, we could establish  that the subdwarf's
companion is a white dwarf. Its X-ray luminosity,  well
constrained by the high quality  \xmm\ spectrum, is much
smaller than that expected from a neutron star accreting in the
stellar wind of \hd . The measurement of the X-ray pulse delays
induced by the orbital motion, and the discovery of the X-ray
eclipse, allowed us to derive the X-ray mass function and to
constrain the system's inclination. This information, coupled
to the accurately measured optical mass function,  gives the
masses of the two stars: M$_{sd}$ = 1.50 $\pm$ 0.05 $\msun$ for
the subdwarf \hd\ and M$_{WD}$ = 1.28$\pm$0.05 $\msun$ for its
white dwarf companion \rx\   \citep{mer09}. The main parameters
of this binary system are summarized in Table \ref{tab-param}.

\rx\ is one of the most massive white dwarfs currently known
and the one with the shortest spin period. This binary  is also
particularly interesting  since systems of this kind, where a
massive white dwarf accretes from a helium star, have been
proposed as possible progenitors of type Ia supernovae
\citep{ibe94,wan09}. Here we report a comprehensive analysis of
all the \xmm\ observations and the results of a search for
optical pulsations.

\begin{table*}[h]
\caption{ Main parameters of the \hr\ binary system
\label{tab-param}
}
\begin{center}
\begin{tabular}{lccc}
 \hline
parameter  & & &  reference$^{(a)}$ \\
\hline
\smallskip
Orbital Period & P$_{orb}$  & 1.5476666$\pm$0.0000022 days &   (1) \\
Eccentricity   & e          & 0                            &  (2) \\
Inclination    & i          & 79--84$^{\circ}$             & (1) \\
Distance       & d          &  650$\pm$100 pc                & (3) \\
   & & & \\
\hline
 \multicolumn{4}{c}{ HD 49798} \\
\hline
Mass                 &  M$_{sd}$ & 1.50$\pm$0.05 $\msun$ &  (1) \\
Effective temperature   &  T$_{eff}$&  46,500 K &  (4) \\
Magnitudes         &     &  U=6.758, B=8.017, V=8.287 & (5) \\
Surface gravity     & log g &   4.35 cgs &  (4) \\
Radius              &   R$_{sd}$   &   1.45$\pm$0.25 $\rsun$ & (3) \\
Luminosity     & L & 3$\times$10$^{37}$  erg s$^{-1}$&(3)\\
   & & & \\
\hline
 \multicolumn{4}{c}{ RX J0648.0--4418 } \\
\hline
Mass   &  M$_{WD}$ & 1.28$\pm$0.05 $\msun$ &  (1) \\
Radius    & R$_{WD}$ & 3000 km &  \\
Luminosity     &  L$_X$&  10$^{32}$  erg s$^{-1}$ &    \\
Spin period   & P & 13.18425$\pm$0.00004  s & (1) \\
Period derivative  & $\pdot$ & --5$\times$10$^{-13}$ s s$^{-1}$ $<\pdot<$ 9$\times$10$^{-13}$ s s$^{-1}$  &   \\
  \hline
\end{tabular}
\end{center}

$^a$ References: (1) \citet{mer09}; (2) \citet{sti94}; (3) \citet{kud78};
(4) \citet{ham10}; (5)  \citet{lan07}.

\end{table*}

\begin{table*}[htbp!]
\caption{Log of the X-ray observations of \hr .
\label{tab-obs}
}
\begin{center}
\begin{tabular}{ccccccc}
\hline \hline
Obs. &  Date    &  start-end              & Exposure     & Orbital &  Spin period \\
     &          & (MJD)                   & pn/MOS (ks)   &  phase  &  (s)     \\
\hline
A & 2002 May 03 &   52397.46--52397.55 & 	4.5 / 7.2      & 0.45--0.48 & 13.18421(7)$^{(a)}$\\
B & 2002 May 04 &   52397.98--52398.06 & 	1.4 / 5.6      & 0.78--0.81 &  --\\
C & 2002 May 04 &   52398.56--52398.59 &	0.6 / 2.5      & 0.15--0.16 & -- \\
D & 2002 Sep 17 &   52534.58--52534.72 &  	6.9 / 11.9     & 0.06--0.12 & 13.1856(14)\\
E & 2008 May 10-11& 54596.90--54597.38 &    36.7 / 43.0    & 0.56--0.87 & 13.18425(4) \\
\hline
\end{tabular}
\end{center}
$^{(a)}$ Joint analysis of observations A, B, and C.
\end{table*}

\begin{table*}[h]
\caption{Results of phase-averaged spectroscopy of the 2008 observation (pn+MOS)
\label{table_spectral}
}
 \begin{center}
 \begin{tabular}{lcc}
 \hline

\smallskip

          &    Blackbody + Power law  &   Blackbody + Bremsstrahlung \\
\hline
\smallskip

N$_H$ (cm$^{-2}$)           & $<$2.4$\times$10$^{19}$        & $<$1.6$\times$10$^{19}$  \\
\smallskip
kT$_{BB}$     (eV)          &   38.9$^{+1.4}_{-1.2}$  &   40.0$^{+1.1}_{-1.3}$\\
\smallskip
R$_{BB}^{(a)}$  (km)        &  17.9$^{+3.2}_{-1.6}$  &  16.3$^{+1.7}_{-1.0}$      \\
\smallskip
$\Gamma$                    &  1.6$\pm$0.1      &     --                        \\
\smallskip
F$_{PL}^{(b)}$ (erg cm$^{-2}$ s$^{-1}$) &  (1.7$\pm$0.1)$\times$10$^{-13}$        & --    \\
\smallskip
kT$_{Brem}$  (keV)    &     --                  &   7.9$^{+2.7}_{-1.7}$         \\
\smallskip
F$_{Brem}^{(c)}$  (erg cm$^{-2}$ s$^{-1}$) &  --       & (1.53$\pm$0.08)$\times$10$^{-13}$    \\
\smallskip
L$_{BB}^{(d)}$ (erg   s$^{-1}$) &    8.9$\times$10$^{31}$    &  8.2$\times$10$^{31}$   \\
\smallskip
kT$_{Ecl}$      (keV)         &  0.55$^{(e)}$  & 0.55$^{(e)}$  \\
\smallskip
F$_{Ecl}^{(c)}$ (erg cm$^{-2}$ s$^{-1}$) & 4.3$\times$10$^{-14}$ $^{(e)}$ & 4.3$\times$10$^{-14}$ $^{(e)}$ \\
\smallskip
$\chi^2$/ dof            &       151.1 / 133          &  158.7 / 133  \\

\hline
\end{tabular}
\end{center}

$^a$ Blackbody emission radius at infinity, for d=650 pc.

$^b$ Observed flux of the power law component 0.2-10 keV.

$^c$ Observed flux of the bremsstrahlung component 0.2-10 keV.

$^d$ Bolometric luminosity of the blackbody component, for d=650 pc.

$^e$ Fixed eclipse component

All the errors are at the 90\% c.l. for a single interesting parameter
\end{table*}


\begin{figure*}
\includegraphics[angle=-90,width=12cm]{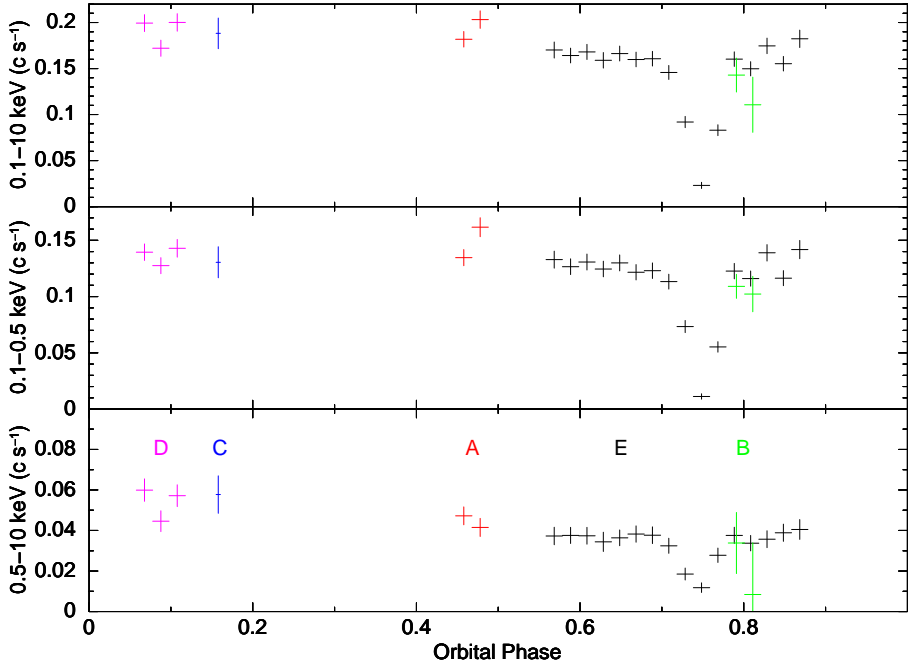}
\caption{ Background-subtracted orbital light curves of \rx\ obtained with the pn camera in three energy ranges: 0.1--10 keV (top), 0.1--0.5 keV (middle), and 0.5--10 keV (bottom).
The labels indicate the five \xmm\ observations listed in Table \ref{tab-obs}.
\label{lcpn} }
\end{figure*}

\begin{figure*}
\label{lcmos}
\includegraphics[angle=-90,width=12cm]{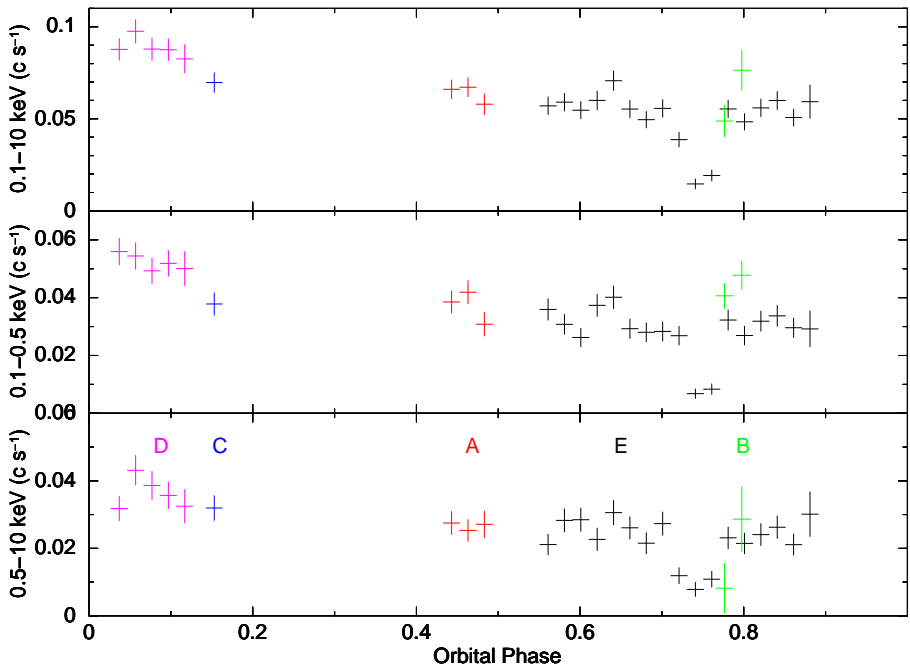}
\caption{ Same as Fig.1,    for the sum of the two  MOS cameras.
\label{lcmos}}
\end{figure*}

\section{X-ray data analysis and results}

\xmm\  performed four short observations of \rx\ in 2002 and a
longer one in 2008 (see Table \ref{tab-obs}). To look for
possible variability, the 2002 pointings were taken at
different orbital phases\footnote{The preliminary results of
the 2002 observations \citep{tie04} are superseded by those
reported here.}. Based on the optical ephemeris of the system
\citep{sti94}, the 2008 observation was scheduled to include
the orbital phase 0.75, in order to look for the presence of an
X-ray eclipse. This orbital phase was not covered by previous
X-ray observations.

Here we concentrate on data obtained with the EPIC instrument (0.1-12 keV), consisting of
two MOS and one pn CCD cameras \citep{tur01,str01}.
During all the observations,   the three cameras were operated in Full Frame mode
(time resolution of 73 ms and 2.6 s for pn and MOS, respectively)
and with the medium optical blocking filter. The data were processed
using SAS version 9.0.

The source photons for the timing analysis were
extracted from a circular region of 30$''$ radius and their arrival
time corrected to the Solar System barycenter and for the orbital motion using the
system parameters given in Table \ref{tab-param}.
The source spectra were   extracted from a slightly smaller
circle (20$''$ radius), in order to avoid possible contamination at
high energy from a source of similar brightness and harder
spectrum located 70$''$ away from \rx . The spectra were rebinned to
have at least 30 counts per energy channel and to oversample by a
factor 3 the instrumental energy resolution. The background spectra
were extracted from nearby regions,
which were on the same chip as the source and where no sources
were detected. Observation B was affected by high particle
background, therefore we excluded it from the spectral
analysis.

\subsection{Variability}

\begin{figure}
\includegraphics[angle=0,width=8cm]{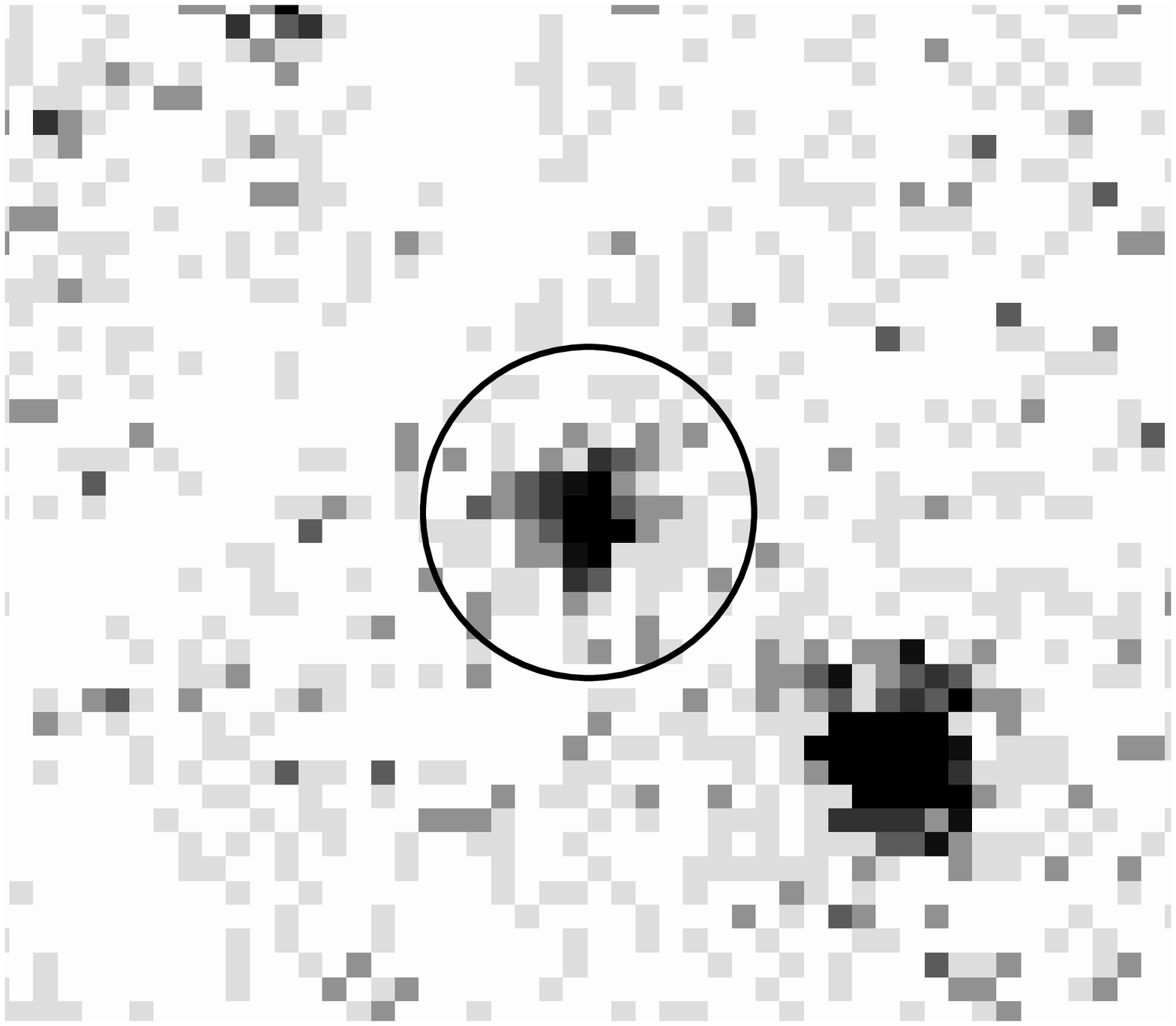}
\caption{X-ray image (0.2--10 keV, pn camera) showing  emission from \hd\
during the orbital phase interval 0.73--0.77, corresponding to the white dwarf eclipse.
The displayed  region has a size of 1.5$\times$1.5 arcmin$^2$; north to the top, east to the left.
The circle, with a radius of 30$''$, indicates \hd .
The source to the SW of \hd\  has a thermal
X-ray spectrum and is probably associated to a late type star of  magnitude V=13.
\label{fig_ima_ecl}}
\end{figure}

The background subtracted light curves of \rx\  in the total
(0.1-10 keV), soft (0.1-0.5 keV), and hard (0.5--10 keV) energy ranges
are plotted as a function of the orbital phase in Fig. \ref{lcpn} (pn)
and Fig. \ref{lcmos} (sum of the two MOS).
These figures clearly show that a significant flux
is detected also during the X-ray eclipse, with net count rates of (2.8$\pm$0.3)$\times$10$^{-2}$
counts s$^{-1}$ in the pn and (1.7$\pm$0.2)$\times$10$^{-2}$ counts s$^{-1}$ in the sum  of the two MOS.
The presence of X-ray emission during the eclipse is confirmed by the
image shown in Fig. \ref{fig_ima_ecl}, which
was accumulated selecting events in the time interval corresponding to the eclipse.

Besides the variation due to the eclipse, the light curves
shown in Figs. \ref{lcpn} and \ref{lcmos} indicate the possible
presence of some variability. In particular, the count rates in
September 2002 (obs.D) were slightly larger than in the 2008
observation, especially at energies above 0.5 keV. This
variability is confirmed by the spectral analysis described
below.

\subsection{Spectral analysis}

We first analyzed  the X-ray emission detected during the eclipse, by extracting the pn and MOS
spectra corresponding to the orbital phase interval 0.73--0.77.
The three spectra were fitted together obtaining good results either with
a power law with photon index  $\Gamma$=2.8$\pm$0.3 or with a thermal bremsstrahlung
with temperature kT$_{Br}$=0.55$^{+0.3}_{-0.2}$ keV ($\chi^2$/dof=17.2/16 and 24.7/16, respectively).
In both cases we fixed the interstellar absorption
to the value N$_H$=2.7$\times$10$^{19}$  cm$^{-2}$ derived from the 2008 un-eclipsed pn data.
The X-rays seen during the  eclipse of the white dwarf are most likely due to  emission from \hd ,
but even in the case of a different origin, we expect that this emission is present at all orbital phases.
Therefore we included its contribution in all the subsequent fits to the non-eclipsed \rx\ emission.

The 2008 out-of eclipse pn and MOS spectra were extracted excluding the orbital
phase interval 0.71--0.79  and fitted jointly with different
models\footnote{With the addition of the eclipse component with parameters fixed at the values of the
bremsstrahlung best fit.}.
Single component models gave unacceptable results, while
good fits could be obtained with two-component models consisting of a blackbody plus either a power law or
a  thermal bremsstrahlung.
The best fit parameters are summarized in Table \ref{table_spectral}.
The results differ slightly from those we reported earlier \citep{mer09}
due to the inclusion in the present analysis of the eclipse component.
The blackbody component, which accounts for most of the flux, is similar in both models,
with temperature kT$_{BB}$$\sim$40 eV and  bolometric luminosity  L$_{BB}\sim$8$\times$10$^{31}$ erg s$^{-1}$
(for d=650 pc).
The hard component provides an additional  luminosity of $\sim$8$\times$10$^{30}$ erg s$^{-1}$
(0.2--10 keV, corrected for the absorption).
To check whether the uncertainties in the eclipse spectrum   affect the overall results,
we repeated the analysis with different bremsstrahlung temperatures
of the fixed component in the range kT$_{Br}$ = 0.3--0.8 keV
and found that the best fit values for the non-eclipse spectra change by less than 10\%.

\begin{figure}
\includegraphics[angle=-90,width=8cm]{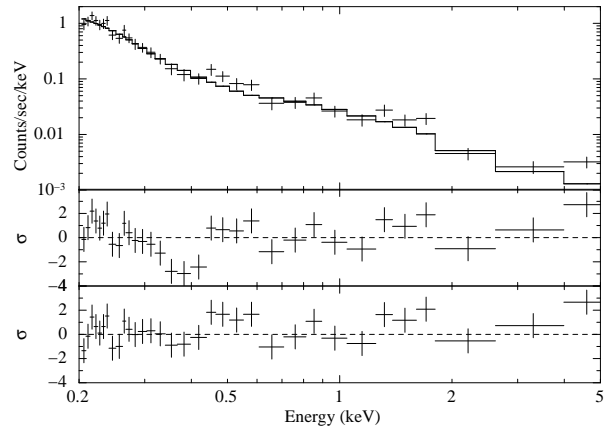}
\caption{Spectrum of \rx\ obtained in May 2002 (Obs. A+C) with the pn camera.
Top panel: data and best fit power-law plus blackbody model (see parameters in Table
\ref{table_spectral_2002}).
Middle panel: residuals obtained with parameters fixed at the values of May 2008 ($\chi^2$/dof = 61.9/32).
Bottom panel: residuals of the best fit model ($\chi^2$/dof = 41.8/32).
\label{fig_spAC}}
\end{figure}

\begin{figure}
\includegraphics[angle=-90,width=8cm]{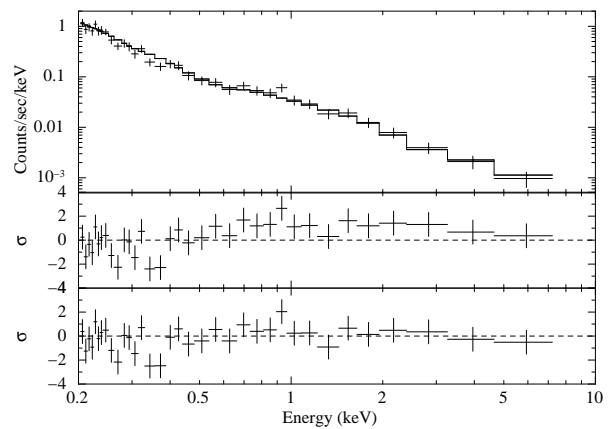}
\caption{Spectrum of \rx\ obtained in September 2002  with the pn camera.
Top panel: data and best fit power-law plus blackbody model (see parameters in Table \ref{table_spectral_2002}).
Middle panel: residuals obtained with parameters fixed at the values of May 2008 ($\chi^2$/dof = 50.7/34).
Bottom panel: residuals of the best fit model ($\chi^2$/dof = 34.7/34).
\label{fig_spD}
}
\end{figure}

Also the spectra of  the 2002 observations rule out single component models,
but, owing to their lower statistics, the best fit parameters have larger uncertainties compared
to those obtained from the 2008 data.
To search for possible long term spectral variations we fitted the 2002
spectra using a  power-law plus blackbody model with photon index,  temperature,
relative normalization, and absorption fixed at the 2008 best fit values.  The   eclipse component
was always included, with all the parameters fixed as described above.
For simplicity we restricted the comparison to the pn data, which are those with the
smallest statistical uncertainties.
The results are shown in Figures \ref{fig_spAC} and \ref{fig_spD},
referring respectively to the observations
of 2002 May (A+C) and 2002 September (D). The residuals obtained in this way (middle panel
of each figure) suggest that in September 2002 only the power law component was brighter than in 2008,
while in May 2002  the largest residuals are in the low-energy part of the spectrum,
where the blackbody component dominates. Allowing the power law normalization to vary in
the September 2002 fit, we obtained  good results  with
a power law flux $\sim$30\% higher than in 2008
(see residuals in the lower panel of Fig.\ref{fig_spD}).
For the May 2002 spectrum, we kept all the power-law
parameters fixed and allowed the blackbody component to change.
This resulted in a best fit temperature kT$_{BB}$=31.6$\pm$0.2 eV and emitting radius
R$_{BB}$=42.5$\pm$1.5 km (see residuals in lower panel of Fig.\ref{fig_spAC}).
Table \ref{table_spectral_2002} summarizes all the parameters obtained
in this analysis of the 2002 observations.

\subsection{Timing analysis}

A best fit pulse period P = 13.18425$\pm$0.00004 s was derived for the 2008 observation by \citet{mer09}.
After correcting the times of arrival to the Solar system barycenter and
for the effects of the orbital motion of the system, we searched for pulsations
in the 2002 observations obtaining the values reported in Table  \ref{tab-obs}.
No long term variations in the spin period could be inferred by comparing the 2002 and 2008
values. A linear fit gives a 90\% c.l. interval   --5$\times$10$^{-13}$ s s$^{-1}$ $<\pdot<$ 9$\times$10$^{-13}$ s s$^{-1}$ for the period derivative.

The folded light curves at soft and hard X-rays are shown in Fig. \ref{fig_profiles}.
No significant variations are seen by comparing the 2002 and 2008 lightcurves.
On the other hand, all datasets show
a clear difference between the two energy ranges. The soft
band, where the blackbody component dominates, has a  nearly sinusoidal
profile with a pulsed fraction of 56\%.
Above 0.5 keV the light curve is instead characterized by two peaks, of unequal intensity
and out of phase with respect to the soft pulse.

\begin{table*}[h]

\caption{Spectroscopy results of the 2002 observations (pn data)
\label{table_spectral_2002}
}
 \begin{center}
 \begin{tabular}{lcc}
 \hline

\smallskip

          &    May 2002 (A+C)  & September 2002 (D)  \\
\hline
\smallskip

N$_H$ (cm$^{-2}$) &   2.7$\times$10$^{19}$$^{(a)}$ & 2.7$\times$10$^{19}$$^{(a)}$    \\
\smallskip
kT$_{BB}$    (eV)  &  31.6$\pm$0.2 &  39.6$^{(a)}$ \\
\smallskip
R$_{BB}$  (km)  & 42.5$\pm$1.5       &  16.6$^{(a)}$   \\
\smallskip
$\Gamma$        &  1.57$^{(a)}$       &   1.57$^{(a)}$             \\
\smallskip
F$_{PL}^{(b)}$  (10$^{-13}$ erg cm$^{-2}$ s$^{-1}$) &  1.7$^{(a)}$   &     2.2$\pm$0.2  \\
\smallskip
L$_{BB}$ (erg   s$^{-1}$) &  2.2$\times$10$^{32}$   & 8.4$\times$10$^{31}$     \\
\smallskip
kT$_{Ecl}$      (keV)         &  0.49$^{(a)}$  & 0.49$^{(a)}$  \\
F$_{Ecl}$  (10$^{-14}$ erg cm$^{-2}$ s$^{-1}$)    &  4.2$^{(a)}$  & 4.2$^{(a)}$  \\
$\chi^2$/ dof            & 41.8 / 31         &  34.7 / 34    \\

\hline
\end{tabular}
\end{center}

$^a$ Fixed


$^b$ Observed flux of the power law component 0.2-10 keV.

All the errors are at the 90\% c.l. for a single interesting parameter
\end{table*}


\subsection{Phase-resolved spectroscopy}

On the basis of the spectral  results  presented above,
clearly showing the presence of two distinct components, we performed phase-resolved spectroscopy
choosing different phase intervals tailored to the light curves seen in the soft and
hard energy ranges. We concentrate here on the 2008 data (excluding the eclipse).
The 2002 data gave fully consistent, but less constrained, results, owing to their smaller statistics.
In all the phase-resolved spectral fits discussed below we included the fixed eclipse component,
as was done for the phase-averaged analysis.

We first searched for spectral variations as a function  of the pulse phase in the soft component,
by  extracting four spectra corresponding to two pairs of phase intervals
selected on the basis of the 0.1--0.5 keV pulse profile:
pulse minimum/maximum (phases 0.25--0.65/0.65--1.25),
and pulse decline/rise (phases 0--0.4/0.4--1).
All the spectra were fitted with the power-law plus blackbody model
keeping N$_H$ linked to a common value.
The  blackbody temperatures obtained in the four spectra were all consistent
with the value kT$_{BB}$$\sim$40 eV found in the phase-averaged spectrum.
Acceptable $\chi^2$ values were obtained by
forcing a common value of kT$_{BB}$ and fitting together the minimum and maximum
or the rise and decline spectra.
Thus we conclude that all the variations in the low energy pulse profiles can be interpreted
as a change in the intensity of the blackbody component that remains at a constant temperature.

\begin{figure*}[ht!]
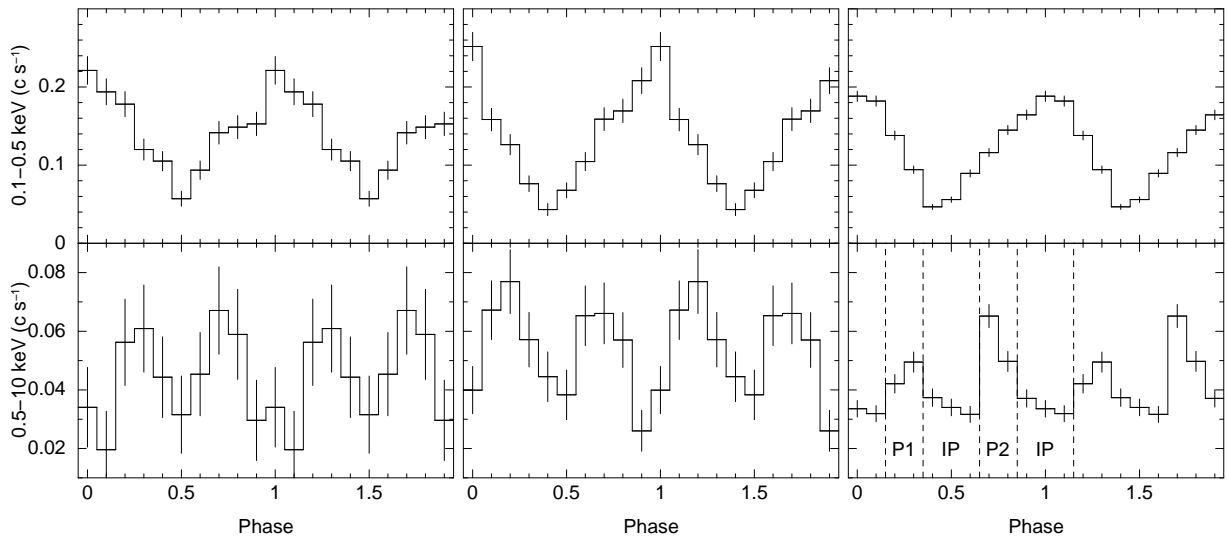

\hbox{
  \includegraphics[angle=-90,totalheight=7cm]{Fig6a.ps}
  \includegraphics[angle=-90,totalheight=7cm]{Fig6b.ps}
  \includegraphics[angle=-90,totalheight=7cm]{Fig6c.ps}
 }
  \caption{ Folded light curves in the soft  and hard energy ranges obtained
  in the observations of May 2002
    (left panel), September 2002 (middle panel) and May 2008 (right panel).
  \label{fig_profiles}  }
  \end{figure*}

\begin{figure}[ht!]
\includegraphics[angle=-90,width=8cm]{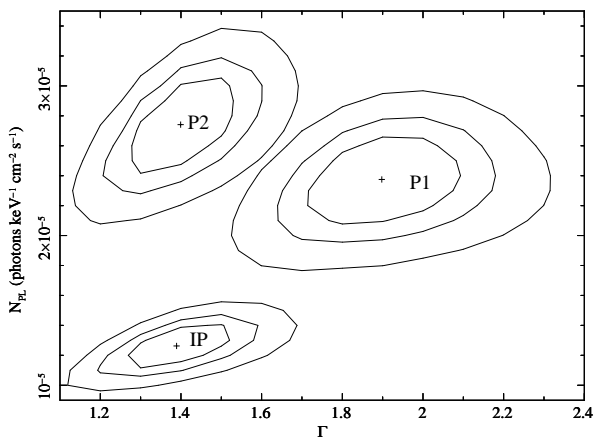}
\caption{Confidence contour of the power law component parameters in phase resolved
spectroscopy. N$_{PL}$ is the flux at 1 keV and $\Gamma$ is the photon index.
The curves correspond to the 68\%, 90\% and 99\% confidence level.
\label{fig_contours}}
\end{figure}

In order to investigate the phase dependence of the hard component,
we extracted  three spectra corresponding to the  intervals marked in Fig. \ref{fig_profiles}:
first peak   (P1, phases 0.15--0.35),
second peak (P2, phases 0.65--0.85),
and inter-peak (IP, phases 0.35--0.65 and 0.85-1.15).
The three spectra were fitted with a blackbody plus power law model, keeping
absorption and temperature fixed to the phase averaged values obtained with the pn
($N_H$=2.7$\times$10$^{19}$ cm$^{-2}$ and kT$_{BB}$=39.6 eV).
The results are given in Table \ref{table_spectral_resolv}.
The spectrum during the first peak is slightly softer than at other phases, as
shown by the confidence level contours for the  power law photon index and normalization plotted in Fig.\ref{fig_contours}.

\section{Search for optical pulsations}

We observed \hr\  with the University of Cape Town CCD photometer \citep{odo95} at the 1.9-m telescope of
the South African Astronomical Observatory in Sutherland on December 25 and 27, 2009.
The observations were carried out in the U filter, with a time resolution of 4 and 6 seconds, respectively.
The first observing run lasted 29.1 minutes, the second run lasted 60.2 minutes.

A timing analysis of the optical data did not reveal any significant signal at, or close to, the frequency of the X-ray pulsations down to a level of 1.5 mmag (December 25), and 0.6 mmag (December 27).
These limits correspond  respectively to a pulsed flux of 1.5$\times$10$^{-3}$ and
0.6$\times$10$^{-3}$  photons cm$^{-2}$ s$^{-1}$ {\AA}$^{-1}$  at  3600 {\AA},
several orders of magnitude above the
extrapolation to the optical U band of the X-ray blackbody spectrum.
Even considering the low energy extrapolation of  the power law  seen during Peak 1,
i.e. the spectral component with the steepest slope ($\Gamma$=1.9$\pm$0.2),
gives a 3600 {\AA} flux of a few 10$^{-6}$ photons cm$^{-2}$ s$^{-1}$ {\AA}$^{-1}$, well below our
upper limit.
However, other mechanisms not directly related to the spectral components seen in X-rays
might in principle contribute in the optical/UV band, and the above limit is the best currently
available in the optical band for this pulsar.

\begin{table}[h]

\caption{Results of phase-resolved spectroscopy of the 2008 observation (pn data)
\label{table_spectral_resolv}
}
 \begin{center}
 \begin{tabular}{lccc}
 \hline

\smallskip

          &    Peak 1  & Peak 2 & Inter-peak \\
\hline
\smallskip

N$_H$$^{(a)}$ (cm$^{-2}$) &   2.7$\times$10$^{19}$ & 2.7$\times$10$^{19}$  &  2.7$\times$10$^{19}$ \\
\smallskip
kT$_{BB}$$^{(a)}$     (eV)  &  39.6 &  39.6 & 39.6 \\
\smallskip
R$_{BB}$  (km)  & 15.1$\pm$0.8       &  17.0$\pm$0.7      &    16.6$\pm$0.4      \\
\smallskip
$\Gamma$        &  1.9$\pm$0.2      &   1.4$\pm$0.15      &      1.39$\pm$0.15             \\
\smallskip
F$_{PL}^{(b)}$  (10$^{-13}$ erg cm$^{-2}$ s$^{-1}$) &  1.55$\pm$0.25  &     2.6$\pm$0.3  & 1.2$\pm$0.2  \\
\smallskip
$\chi^2$/ dof            & 38.4 / 28         &  43.8 / 32     &  83.7 / 62   \\

\hline
\end{tabular}
\end{center}

$^a$ Fixed

$^b$ Observed flux of the power law component 0.2-10 keV.

All the errors are at the 90\% c.l. for a single interesting parameter
\end{table}


\section{The X-ray emission from \rx\ }

The results of X-ray observations with \xmm\ spanning six years indicate that the properties
of \rx\ are rather stable. Only minor changes were seen in the spectrum, but it is not
possible at this stage to attribute them to the different orbital phases of the observations or
to a long term variability. The spectrum and luminosity
seen with \xmm\ are  consistent with those measured with \textit{ROSAT} in 1992
and \textit{Einstein} in 1979, although the uncertainties in the latter data are rather large.
No significant variations were seen in the spin period, pulsed fraction and shape of the pulse profile.
Given that this is the only known X-ray emitting binary composed of a hot sub-dwarf and a white dwarf,
it is worth examining different possibilities for the origin of the observed X-ray emission.

The upper limit on the white dwarf secular spin-down ($\pdot$ $<$9$\times$10$^{-13}$ s s$^{-1}$),
implies  a rotational energy loss I$_{WD}$ 4$\pi^{2}$ P$^{-3}$ $\pdot$
\ltsima1.5$\times$10$^{36}$ erg s$^{-1}$
(for a moment of inertia I$_{WD}$ = 10$^{50}$ g cm$^{2}$).
This is sufficiently high to power the observed    luminosity
of $\sim$10$^{32}$ (d/650 pc)$^2$ erg s$^{-1}$, with an efficiency similar to that
seen in rotation-powered neutron stars.
Even if no convincing examples have been discovered, the existence of white dwarf analogues
of rotation-powered neutron stars is an intriguing possibility (see, e.g., \citet{uso93}).
Among isolated white dwarfs, even the one with the fastest rotational velocity
(RE J0317--853, P=726 s, \cite{ven03})
can develop a maximum  potential drop along open
field lines\footnote{Assuming a dipole field} of only $\Delta$V = $\Omega^2$ B R$^3$ / c$^2$
$\sim$7$\times$10$^{9}$ (726 s/P)$^2$ (B/660 MG) (R/5000 km)$^3$ V, despite its high magnetic field (B$\sim$170--660 MG).
This should be compared with  the
much higher value $\sim$10$^{15}$--10$^{17}$ V typically reached in radio pulsars.
White dwarfs in binary systems can rotate much more rapidly.
After \rx , the next fastest rotators are found in the dwarf nova WZ Sge
(P=28.87 s, \cite{pat98}) and in the intermediate polar AE Aqr (P=33 s, \cite{pat79}).
Their magnetospheres can, in principle, accelerate particles to high energies, but still few
orders of magnitude below that reached in radio pulsars.
A value  $\Delta$V $\sim$10$^{14}$ V could be attained in \rx , but
only  in the case of a strong magnetic field, which, as we discuss below, seems rather unlikely.
The possible detection of pulsed hard X-rays (10--30 keV) in AE Aqr has been interpreted
as evidence for non-thermal processes powered by the white dwarf rotational energy \citep{ter08},
implying  an efficiency of $\sim$0.1\%.  AE Aqr spins-down at a rate of 5.6$\times$10$^{-14}$
s s$^{-1}$, most likely as a result of the  magnetic propeller effect.
Its K-type main sequence companion transfers mass trough Roche-lobe overflow, but the
accretion stream is disrupted by the magnetic field of the white dwarf and most of the
mass is ejected \citep{wyn97}.
AE Aqr exhibits a hard X--spectrum and strong rapid variability at all wavelengths,
contrary to \rx\ which shows a nearly constant flux, dominated  by a thermal-like component.
The significant differences, except for the short spin-period, between these systems disfavor
the presence of a rotation-powered mechanisms in \rx .

We believe that accretion provides a much more plausible and natural explanation for the  X-ray emission
observed from \rx .
The Roche lobe of \hd\ has a radius R$_L$ $\sim$  a (0.38+0.2 Log (M$_{sd}$/M$_{WD}$)) = 3.2 $\rsun$,
where a=5.5$\times$10$^{11}$ cm is the orbital separation.
The sub-dwarf, with a radius R$_{sd}$=(1.45$\pm$0.25) $\rsun$  \citep{kud78}, is significantly smaller
than its Roche-lobe. Therefore, accretion can only occur via stellar wind capture.
Hot sub-dwarfs have relatively weak stellar winds, compared to  main sequence or
giant stars of the same spectral type, but \hd\ is one of the few sdO stars for which evidence of
a significant mass loss has been reported.
The P Cygni profiles of its N V and C IV lines,  indicate
a mass loss in the range  5$\times$10$^{-10}$--10$^{-8}$ $\msun$ yr$^{-1}$
and a wind terminal velocity v=1350 km s$^{-1}$,
reached at only 1.7 R$_{sd}$ \citep{ham81}.
A more precise estimate of $\Mdot$ = 3$\times$10$^{-9}$ $\msun$ yr$^{-1}$ has been obtained thanks to a
recent analysis of the optical/UV spectra based on improved wind models \citep{ham10}.
The accretion rate onto the white dwarf can be estimated
as $\Mdot_{WD}$$\sim$ $\Mdot$ (R$_a$/2a)$^2$ = 8$\times$10$^{-13}$ $\msun$ yr$^{-1}$, where
R$_a$$\sim$2GM$_{WD}$/v$^2$ is the accretion radius.
The corresponding accretion luminosity is 3$\times$10$^{31}$ (R$_{WD}$/3000 km)$^{-1}$ erg s$^{-1}$,  in good agreement with the observed value.

The presence of two components in the X-ray spectrum of  \rx\ is similar to what is observed
in many  cataclysmic variables of the polar and intermediate polar classes.
In these systems, the magnetic field starts to dominate the motion of the accreting matter at
large distances from the white dwarf, channeling the flow onto the star's magnetic pole(s),
and even preventing the formation of an accretion disk in the systems
with the strongest fields.
Shocks are formed in the accretion column above the white dwarf and the hot plasma in
the post-shock region emits hard X-rays. This hard component is usually fit with
multi-temperature plasma models (see, e.g., \citet{cro99,yua10}), but, as a first approximation, it
can  be described by a thermal bremsstrahlung with temperature kT$\sim$30 (M/$\msun$) R$_9^{-1}$ keV,
where M is the white dwarf mass and R$_9$ its radius in units of 10$^9$ cm.
Part of the hard X-ray emission may be intercepted and reprocessed by the white dwarf surface,
giving rise to UV and soft X-ray thermal emission, with blackbody temperature of tens of eV.
The relative importance of the soft and hard component depends on
several factors, including geometrical effects, the white dwarf mass,
magnetic field, accretion rate per unit surface, with the consequence that
the observed values of the ratio $L_{soft}/L_{hard}$ span a large range \citep{eva07}.

The value $L_{soft}/L_{hard}$ $\sim$10 observed for  \rx\ is
rather high, but not uncommon. For example, similar or larger
values are found in  EU UMa, DP Leo, and RX J1007--2016
\citep{ram04}. The blackbody temperature we derived for the
soft component, as well as the dimensions of the emitting
surface are well within the range of values measured for the
polars. On the other hand, the  best fit temperature of the
bremsstrahlung component, kT$_{TB}\sim$8 keV is rather small if
one considers that, based on the relation mentioned above,
massive white dwarfs should have the highest shock
temperatures. This discrepancy might have to do with the unique
properties of this white dwarf, which accretes from a low
density stellar wind rather than through Roche-lobe overflow.
The    intensity and configuration of the magnetic field is
also relevant for the properties of the X-ray emission, and
might explain the differences between this system and classical
cataclysmic variables. An upper limit on the dipolar field of
\rx\ can be set by the requirement that its magnetospheric
radius, R$_M$,  be smaller than the corotation radius R$_{COR}$
= (G M$_{WD}$ P$^{2}$/4$\pi^{2}$)$^{1/3}$ = 9$\times$10$^{8}$
cm. Since R$_M$ is defined by balancing the magnetic pressure
and the ram pressure of the accreting matter, this translates
in an upper limit on the white dwarf magnetic moment
$\mu\sim$2$\times$10$^{29}$ G cm$^3$. However, the above
condition does not exclude the possibility that the source be
in the sub-sonic propeller regime, in which  R$_M$$<$R$_{COR}$,
but the matter is too hot to accrete \citep{dav79}. Using the
equation that defines the transition between the accretion
regime and the subsonic propeller regime \citep{boz08}, we
obtain  an upper limit $\mu$$<$3 10$^{28}$ G cm$^3$,
corresponding to  a field of  the order of  1 kG at the white
dwarf's surface.

\section{\hd :  the first X-ray emitting hot sub-dwarf}

Early type stars emit X-rays with relatively soft thermal
spectra (kT$\sim$0.5 keV) and luminosity proportional to the
bolometric luminosity. The average relation L$_X$
$\sim$10$^{-7}$ L$_{bol}$ has been derived from the
observations of  main sequence, giant and supergiant stars of O
spectral type. The X-ray emission is believed to originate from
hot plasma heated by  shocks and instabilities in the strong
stellar winds of these stars  \citep{pal89}. Despite their very
high effective temperature, O type sub-dwarfs are much fainter
than main sequence stars, and, up to now, X-ray observations
provided only an upper limit of $\sim$10$^{31}$ erg s$^{-1}$
for the sdO star BD --3$^{\circ}$2179 \citep{dan83}.

The X-rays we detected during the  eclipse are well described with a   bremsstrahlung
of temperature 0.5 keV and luminosity 2$\times$10$^{30}$ erg s$^{-1}$.   Given the bolometric
luminosity of \hd\ L$_{bol}$=10$^{3.9}$$\lsun$, this corresponds to a ratio
L$_X$/L$_{bol}$ = 7$\times$10$^{-8}$. Thus, both from the spectral and luminosity point
of view, the observed X-ray flux is consistent with emission from \hd .
To our knowledge, this could be the first detection of X-ray emission from a sdO star.
Further observations of this system, as well as searches for X-ray emission from other
members of this class, can provide interesting information on the properties of X-ray production
in stellar winds spanning a large range of parameters, from hot sub-dwarfs to supergiants.

\section{Discussion}

\hr\  is the only known X-ray  binary containing a massive white dwarf accreting from a hot sub-dwarf.
In principle, other systems of this kind, with a luminosity of $\sim$10$^{32}$ erg s$^{-1}$,
could  be detected even at distances of several kpc by the current X-ray satellites.
However, due to the  very soft spectrum, the observed flux
is very sensitive on the amount of interstellar absorption.
For example, a column density of 3$\times$10$^{21}$ cm$^{-2}$ would reduce by one order of magnitude
the observed EPIC count rate of \rx , making it
difficult to identify other sources of this kind in the Galactic plane only through X-ray observations
(\rx\ is at  Galactic coordinates l=254$^{\circ}$, b=--19$^{\circ}$).

On the other hand, also from the point of view of the optical properties, this system seems quite uncommon.
In fact,  \hd\ is rather atypical compared to  other sdO stars \citep{heb09}, which on average are
less massive (M$\sim$0.5 $\msun$),  less luminous, and have a higher surface gravity.
This is probably  a result of its evolution in a close binary in which a common envelope
phase occurred, as clearly indicated by several arguments.
The low H abundance (X$_H$$\sim$0.19, \citet{kud78}),
suggests that \hd\ is the stripped core of an initially much more massive and larger star.
Also the high abundance of N and low abundance of C confirm that its
present surface layers once belonged to the outer part of
the hydrogen-burning core of a massive star.
Slightly different evolutionary scenarios  have been proposed to explain this system.

One possibility is that \hd\ descend from a fairly massive progenitor that began to fill its Roche
lobe before helium ignition (Case B evolution) and is currently in
a core He-burning phase \citep{kud78,ibe93}.  In this case,  the mass of the He star
is related to that of its progenitor by M$_{He}$=0.043 M$^{1.67}$ \citep{ibe85},
implying that \hd\ was originally a star of $\sim$8 $\msun$.
Two other possibilities have been considered by  \cite{bis97}.
The first one is that \hd\ consist of a degenerate CO core,
surrounded by a helium envelope with a He-burning shell at its base. In this scenario,
it would be the core of a progenitor of $\sim$5 $\msun$ which lost mass during a
common envelope event when it was on the early asymptotic giant branch (AGB).
Alternatively, \hd\ could be burning  hydrogen in a shell, being the remnant
of a star which shed its envelope while on the thermally pulsing AGB. However,
this possibility is disfavored since it predicts a  mass of only $\sim$0.6 $\msun$
for the sub-dwarf.

At the observed mass transfer rate, with only a small ($<10^{-3}$) fraction of the total
mass lost in the subdwarf's wind accreted, the white dwarf mass is increasing very slowly.
At the end of the current He-burning phase, \hd\ will expand and fill the Roche-lobe, giving
rise to a much higher mass transfer rate. This will produce a higher X-ray luminosity and possibly
bright outbursts like the one seen in the helium nova V445 Puppis \citep{wou09},
which could be the descendent of a system similar to \hr .
The amount of accreted mass,  $\sim$0.3--0.5 $\msun$  (see figure 1 of \cite{ibe94}),
is sufficient to bring the, already quite heavy, white dwarf above the Chandrasekhar limit.
However,  the fraction of mass that is effectively retained  depends on the rate and composition of the accreting matter, on the mass, composition
and temperature of the white dwarf, and on the poorly known relevance of other factors,
such as, e.g., rotation, magnetic fields, wind outflows. Thus the fate of \rx\ is uncertain.

Evolutionary computations for \rx\ have been recently performed \citep{wan10}
assuming the mass accumulation efficiency  that takes into account the wind mass loss
triggered by the He-shell flashes  \citep{kat04}.
These  indicate that \rx\ will reach a mass of 1.4 $\msun$ after only a few
10$^4$ years of Roche-lobe overflow, during which $\sim$(5--6)$\times$10$^{-6}$ $\msun$ yr$^{-1}$ of
He-rich matter are steadily converted to C and O, while the unburned matter is ejected by the system
trough an optically thick wind.
\cite {wan10} assumed that \rx\ is a CO white dwarf that will explode as a type Ia supernova
when the Chandrasekhar limit is reached. The fast rotation can increase the mass stability limit
above the standard value for non-rotating stars, thus systems like \rx\ could be the progenitor
of over-luminous type Ia supernovae.
While an ONe composition seems more likely for \rx\ in view of its large mass,
also in this case  there are some uncertainties and other factors that might play a crucial role.
For example, it has been pointed out that massive white dwarfs can have a CO composition as
a result of rapid  rotation \citep{dom93,dom96}.

If \rx\ is an ONe white dwarf, an accretion induced collapse (AIC) might occur, instead than a type Ia supernova explosion, 
leading to the formation of a millisecond pulsar.
There is strong evidence that  MSP are old neutron stars which underwent  magnetic field decay
and spin-up due to  accretion of mass and angular momentum in low mass X-ray binary systems (LMXRB) \citep{bha91}.
This  evolutionary scenario is supported by the observation of short
period X--ray pulsations in the persistent
and/or burst emission of several LMXRB, by the small eccentricity of  MSP
with white dwarf companions in the Galactic disk\footnote{This is due to tidal circularization
during the long LMXRB phase; eccentric MSP binaries can be formed if the companion
is another neutron star that received a kick in the supernova explosion in which it was born,
or for systems formed in dense environments where stellar dynamic interactions are important,
like globular clusters.},
and by the recent discovery of PSR J1023+0038 \citep{arc09},
the long sought missing-link between the LMXRB and MSP phase.
On the other hand, the formation of MSP by AIC of accreting white dwarfs
(e.g., \cite{bai90}) has been considered to reconcile the
apparent discrepancy in the birthrate of LMXRB and MSP,
and it cannot be excluded that   a fraction of the MSP population is formed through AIC.
The discovery of PSR J1903+0327, a Galactic plane MSP with a high eccentricity orbit (e=0.44)  around
a solar mass main sequence star \citep{cha08,fre11}, gives further evidence for the
possibility of directly forming young MSP.
The high spin rate and low magnetic field of \rx\ make it an ideal candidate for the  direct
formation of a MSP though AIC.

\section{Conclusions}

\hr\ is a unique system, but its X-ray properties are in several respects
similar to those seen in white dwarfs of the polar and intermediate polar class.
Its spectrum consists of a very soft, strongly pulsed, thermal like emission
plus a harder component dominating above $\sim$1 keV. The two components
have different pulse profiles and show small uncorrelated variations on long time scales.
These similarities suggest that the X-ray emission properties depend mainly on the
physical conditions near the white dwarf, with little influence of the large scale accretion scenario
(wind accretion in \rx\ wrt Roche lobe overflow in cataclysmic variables).

With a dynamically measured mass of 1.28$\pm$0.05 $\msun$, \rx\ is one of the most massive white
dwarf currently known and the one with the shortest spin period,
only a factor $\sim$5 larger than the break-up limit.
It is also the only known white dwarf accreting from the wind of a hot sub-dwarf star.
This system is the outcome of a common envelope evolution, most likely of an original pair of
stars with mass of $\sim$8--10 $\msun$.  Its future evolution might lead to the formation
of a neutron star through accretion induced collapse or to the explosion of a type Ia supernova.

Both  possibilities have interesting implications. Accretion
induced collapse of a fast-spinning and low magnetic field
white dwarf could be a promising scenario for the direct
formation of non-recycled millisecond pulsars. In the second
case, being the result of the evolution of relatively massive
stars ($\sim$8--9 $\msun$), this would be a type Ia supernova
formation channel with a short delay time. Future X-ray and
optical observations can lead to more accurate determination of
the system parameters, providing crucial information to
investigate in more detail the past and future evolution of
this system. For example, further observations of the orbital
phases including the eclipse can give a more precise
measurement of its duration, a better characterization of the
X-ray emission from \hd , as well as some information on the
structure of its wind by measuring possible variations of
N$_H$. High-resolution spectroscopy of the soft X-ray spectral
component and deeper observations above a few keV are needed to
better compare the properties of this peculiar X-ray binary
with those of other accreting white dwarfs, and, in particular,
to better constrain the magnetic field geometry and intensity.

\acknowledgements

We thank Alak Ray, Ulrich Heber and Stephan Geier for
interesting discussions. We acknowledge financial contribution
from the agreements ASI-INAF I/009/10/0 and I/032/10/0. PE
acknowledges financial support from the Autonomous Region of
Sardinia through a research grant under the program PO Sardegna
FSE 2007--2013, L.R. 7/2007 ``Promoting scientific research and
innovation technology in Sardinia''.


\end{document}